%% file: vikram.tex
\documentclass[useAMS,usenatbib,usegraphicx]{mn2e}

\title[Young SNRs Hadronic Emission ]{Exploring the Gamma-Ray Emissivity of Young Supernova Remnants I: Hadronic Emission}
\author[Dwarkadas]{V. V. Dwarkadas\thanks{E-mail:
    vikram@oddjob.uchicago.edu} \\ Department of Astronomy and Astrophysics, U Chicago, 5640 S Ellis Ave, Chicago, IL 60637}
  
\begin{document}
\input{macros}

\date{}

\pagerange{\pageref{firstpage}--\pageref{lastpage}} \pubyear{2013}

\maketitle

\label{firstpage}

\begin{abstract}
Using a simplified model for the hadronic emission from young
supernova remnants (SNRs), we derive an expression to calculate the
hadronic luminosity with time, depending on the SN ejecta density
profile and the density structure of the surrounding medium. Our
analysis shows that the hadronic emission will decrease with time for
core-collapse SNe expanding in the winds of their progenitor stars,
but increase with time for SNe expanding into a constant density
medium, typical of Type Ia SNe. Using our expressions, we can compute
the time-dependent hadronic flux from some well-known young SNe and
SNRs with time, and where applicable reproduce previous results in the
appropriate parameter regime. Using our calculations, we also
emphasize the exciting possibility that SN 1987A may become a visible
gamma-ray source in the next decade.
\end{abstract}

\section{Introduction}
\label{sec:intro}

Supernova remnants (SNRs) are considered as a source of very high
energy accelerated particles, at least up to the knee of the
cosmic-ray spectrum.  Details of the process by which the acceleration
happens are not completely understood, but it is assumed to be related
to Diffusive Shock Acceleration \citep[DSA,][]{drury83, md01} and its
nonlinear modification \citep{edm97}. The discovery of X-ray
synchrotron emission from SN 1006 provided the first convincing
evidence that electrons can be accelerated to TeV energies. Recent
discoveries of SNRs at $\gamma$-ray wavelengths have further supported
the notion that SNRs can accelerate particles, both electrons and
protons, to GeV and even TeV energies. Most recently, direct detection
of the pion decay signature in 2 SNRs conclusively shows the presence
of accelerated protons in SNRs \citep{fermipion13}.

The success of the {\it Fermi} telescope, coupled with ground-based
Cerenkov telescope arrays, has considerably increased the number of
supernova remnants (SNRs) observed in $\gamma$-rays. At present the
number is large enough that one can begin to determine statistical
properties of SNRs, and correlations between their multiwavelength
properties. The upcoming release of the Fermi-LAT Supernova Catalog
will allow for a comprehensive study of SNRs in gamma-rays, and
comparison with multi-wavelength data. The purpose of the catalog is
to systematically investigate the properties of observed gamma-ray
SNRs in a multi-wavelength context \citep{ba12}. It will characterize
GeV emission from SNRs, examine multi-wavelength correlations,
determine statistically significant SNR correlations, and calculate a
spectral model. {\em The existence of a large number, the continual
  detection of many more SNRs, and the very nice cataloging of their
  high energy and multi-wavelength properties, requires a commensurate
  investment of directed theoretical effort to fully understand the
  multi-wavelength and high energy properties of SNRs.}

In the past decade, there has been considerable interest in
understanding the very high energy emission from, and broadband
spectrum of, SNRs. There exist very good papers by many groups that
carry out sophisticated models for understanding the broadband
multiwavelength emission from SNRs, and in particular determining
whether the very-high energy (GeV-TeV) emission is due to hadronic or
leptonic processes \citep{bv10, ekjm11, caprioli11, espb12, csep12,
  ad12, mc12, bkv13}. Such modelling often requires very complicated
codes containing a variety of different physics, which have been
developed mainly in the last decade. However, even with sophisticated
modelling, it has been hard to understand the origin of the very high
energy emission in most SNRs, and to delineate the leptonic and
hadronic contributions.  Unfortunately, complicated models that
require several weeks to months of work per SNR are not conducive to
understanding a statistical ensemble of SNRs. What is needed is a
simpler, but still reasonable accurate way to understand the
statistical properties of SNRs, that can reasonably reproduce the
spread in overall properties even if it cannot accurately fit the
emission from individual SNRs. Such models must necessarily be
(semi)-analytic (although informed by the results of numerical
calculations), where the effect of varying one or more parameters,
keeping all else constant, and studying the results on a statistical
sample of SNe, can be achieved in a short time.

In anticipation of the need to understand the general properties of an
ensemble of SNRs, and the release of the Fermi SNR catalog, we have
embarked on a project to explore the properties of gamma-ray SNRs,
especially young SNRs that have not reached the Sedov-Taylor or
adiabatic stage. Using semi-analytic arguments coupled with realistic
approximations of SNR evolution, we plan to study the gamma-ray
emissivity of SNRs, and investigate the time evolution of the
gamma-ray luminosity due to pion decay and leptonic processes
(non-thermal bremsstrahlung and Inverse Compton emission). In this
first paper we concentrate on the hadronic emission due to pion
decay. We show how the gamma-ray visibility is related to various SN
properties, as well as the properties of the surrounding medium. We
explore various types of SNe and show how their expansion in different
environments affects their temporal evolution. Our calculations allow
us to map out a range of luminosities for various SN types.

The paper proceeds as follows: In \S 2 we outline the evolution of
young SNRs (those which have not yet reached the Sedov stage), and
reasons why it is important to study them. In \S 3 we compute an
analytic expression for the luminosity of young SNRs due to pion
decay, that illustrates the various factors that affect the
luminosity. We also explore expressions for the luminosity in
different environments. In \S 4 we apply the model to individual SNe
and SNRs such as Cas A, SN 1993J and SN 1987A. \S 5 puts our work in
the context of previously published literature on the subject, and
discusses its successes, shortcomings and implications. Finally, \S 6
summarizes our work, and outlines future directions.

\section{Evolution of Young Supernova Remnants}
\label{sec:snrevol}

In the past, the young SNR phase, which we alternately refer to here
as the ejecta-dominated phase, has often been neglected in
considerations of the particle-acceleration process. This is slowly
changing in recent times \citep{pzs10, caprioli11, eb11, espb12,
  tdp12a, tdp12b, zp12, tdp13a}. The reasons generally given for
neglecting it are that the phase only lasts for a short time, until
the swept-up mass equals the ejecta mass; the shock velocity is
constant \citep[e.g.][]{pz03, ga07, np12}; and that the maximum energy
of the particles increases up until the end of the ejecta-dominated
stage and the beginning of the Sedov-Taylor stage
\citep{helderetal12}, when it is that particles start
escaping. However, all of these assertions are false
\citep{dwarkadas11a}. As has been shown by many authors, since at
least \citet{gull73}, the time taken to reach the Sedov-Taylor stage
is much larger than that assumed in many papers. It requires the
swept-up mass to significantly exceed the ejecta mass, by a factor of
20-30, depending on the ejecta profile \citep{dc98}. This considerably
increases the time taken to reach the Sedov stage, with the result
that a 1000 year old remnant such as SN 1006 is still very far from
the Sedov stage. Similarly, SNRs expanding in the low-density
interiors of wind-blown bubbles from massive stars may take a long
time to sweep-up enough mass to reach the Sedov stage, and then in
some cases may bypass the Sedov stage altogether
\citep{Dwarkadas2005}. The second assertion, that the velocity is
approximately constant, is also incorrect. The velocity is continually
decreasing in the ejecta dominated stage. Although this is shown by
several calculations \citep{chevalier1982a}, the definitive proof is
in the observations of young SNRs such as Cas A, Tycho and SN 1006,
where the observed shock velocities are much lower than the fiducial
value of $ > 10^9$ cm s$^{-1}$ that has been conveniently assumed as
the ``free-expansion'' velocity in many papers. The last assertion,
that the maximum energy of accelerated particles is only reached at
the beginning of the Sedov stage, has been disproved by several
arguments \citep{vt09, dtp12b} as well as excellent numerical
simulations \citep{bsrg13}. They show that the maximum energy is
reached early in the young SNR stage in most cases, after which it
begins to decrease. To summarize, the young ejecta-dominated SNR stage
lasts for a long time, potentially up to several thousand years for
SNRs exploding in a low density region or wind-blown cavity. The
maximum energies to which particles are accelerated is reached
generally early in this stage, the velocity continually decreases, and
the escape of particles from the SNR happens in this phase. Given
these considerations, we feel that it is very important to study SNRs
in this stage.

In order to describe the evolution of a young SNR, we use the
formulation suggested by Chevalier
\citep{chevalier1982a,chevalier1994}. In brief, the expansion of SN
ejecta into the surrounding medium leads to the formation of a
double-shocked structure, consisting of a reverse shock that travels
back into the ejecta, and a forward shock that expands into the
ambient medium. From analytical models as well as hydrodynamic
simulations of exploding core-collapse stars, it has been found that
SN ejecta can be effectively described by ${\rho}_{ej} =
A\,v^{-n}\,t^{-3}$, where the coefficient $A$ depends on the explosion
energy and the mass of the ejecta, and can be evaluated with the
information given in \citet{chevalier1994}. The surrounding medium can
generally be ascribed as having a power-law profile in density,
${\rho}_{amb} =B\,r^{-s}$, where $s=0$ denotes a constant density
medium, and $s=2$ denotes a wind medium with constant mass-loss rate
$\dot{M}$ and constant wind velocity $v_w$, with the coefficient $B=
\dot{M}/(4 \pi v_w)$. The resulting expansion of the contact
discontinuity happens in a self-similar fashion, with the self-similar
solutions given by \citep{chevalier1982a}:

\be
R_{CD} = \left(\frac{\delta A}{B}\right)^{1/(n-s)} t^{(n-3)/(n-s)}
\ee

\noindent
where the value of the parameter $\delta$ is given in
\citet{chevalier1982a}.  We can write the expansion of the contact
discontinuity as R$_{CD} = C_1 t^m$, where $m=(n-3)/(n-s)$ is referred
to as the expansion parameter. Note that since the solutions require
$n > 5$, and $s < 3$, we have $m < 1$. Since the expansion is
self-similar, the forward and reverse shocks will expand in the same
manner. In the self-similar case the ratio of the shocks to the
contact discontinuity, and to each other, will be fixed. We can write
the radius of the forward shock as $R_{sh} = \kappa R_{CD} = \kappa
C_1 t^m$. The velocity $v_{sh} = d\,R_{sh}/dt = m \kappa C_1 t^{m-1}$,
and is therefore always decreasing with time.

\section{Hadronic Emission From Young Supernova Remnants}
\label{sec:hadron}

The interaction of protons accelerated at the SNR shock front with
protons in the interstellar medium gives rise to neutral pions (among
other species) which subsequently decay to give gamma-rays. The
gamma-ray flux of SNRs due to hadronic emission can be written as
\citep[][hereafter DAV94]{dav94}\footnote{In principle the hadronic
  flux needs to be written as $$ \frac{q_{\gamma}}{4 \pi d^2}\,\int n
  \left[\frac{{\epsilon}_{CR}}{V}\right] 4 \pi r^2 dr $$ where $n$ is
  the density of the surrounding medium. This gives the same
  time-dependence as found here, with some variation on order unity in
  the normalization factor. We prefer to write it this way so that we
  can easily differentiate the contribution of various shocks, and
  delineate various contributing factors, as shown in this paper.}:

\be
F_{\gamma} (> E_o,t) =  \frac{q_{\gamma}}{4 \pi d^2}\, \frac{M(t)}{\mu m_p}\, \left[\frac{{\epsilon}_{CR}}{V}\right]
\label{eq:pion}
\ee

\noi where $q_{\gamma}$ is the $\gamma$-ray emissivity normalized to
the cosmic-ray energy density, and tabulated in \citet{dav94}. Note
that all the information regarding the nuclear interactions and
spectrum of accelerated particles is contained in the parameter
$q_{\gamma}$.  ${\epsilon}_{CR}$ is the energy in cosmic rays, $V$ is
the emitting volume and $d$ is the distance to the source. $M(t)$ is
the mass of material with which the accelerated protons are
interacting, $\mu$ is the mean molecular weight, and $m_p$ the proton
mass.

We consider a SNR expanding in a medium with surrounding density $B
r^{-s}$, where $B$ is assumed to be a constant. As the SN shock
expands outward at high velocity, it sweeps up the material ahead of
it. The mass swept-up by the shock, which forms the ``target'' mass
$M(t)$ with which accelerated protons will interact to give hadronic
emission, is given by:

\be
M_{sw} = \int^R_0 4 \pi r^2 B r^{-s} dr = \frac{4 \pi B}{3-s}\,R^{3-s}
\ee

The energy density of cosmic-rays from SNRs is a highly-debated
question. Here we take the approach that a fraction of the energy
available at the shock front is used to accelerate cosmic-rays.  The
maximum energy that can be extracted from the shock front at any given
time in the ejecta-dominated phase is less than the total kinetic
energy of the explosion, and is $2 \pi R_{sh}^2 {\rho}_{sh}
V_{sh}^3$. The total energy expended in cosmic ray acceleration up to
a given time $t$ will be some fraction of the integral of this
quantity over time $t$:

\begin{eqnarray}
{\epsilon}_{CR} & = & 2 \pi \int_{0}^{t} \xi R_{sh}^2 {\rho}_{sh} V_{sh}^3\; dt\\
&  = & 2 \pi B \xi {(\kappa C_1)}^{5-s} m^3 \int_{0}^{t} t^{2m-ms+3m -3}\; dt \\
& = & \frac{2 \pi B \xi {(\kappa C_1)}^{5-s} m^3}{5m-ms-2} t^{5m-ms -2}
\label{eq:maxen}
\end{eqnarray}

\noi where $\xi$ denotes the fraction of the shock energy that is
converted to cosmic rays. For convenience we assume here that $\xi$ is
a constant, but there is no reason why it should be so, and it is
quite possible that it could be a function of time. We note that for a
constant density medium ($s=0$), the available energy at the shock
front increases with time as long as $m> 2/5$, i.e. as long as the SNR
has not entered the Sedov-Taylor stage, when all the kinetic energy is
available. Similarly, for a wind medium, the available energy goes as
$t^{3m-2}$ and thus will continually increase until $ m=2/3$, which is
when the remnant enters the Sedov-Taylor phase in a wind ($r^{-2}$)
medium with constant wind parameters.

The volume of the shocked region (the most likely to provide target
protons) can be written approximately as $V = \beta 4 \pi R_{sh}^3/3
$.  Here $\beta \sim 0.3-0.5 $ accounts for the fact that the volume
of the shocked region from which the emission arises, between the
forward and reverse shocks, is smaller than the volume of the entire
SNR.

Putting all the above back in eqn \ref{eq:pion}, we get:

\be
F_{\gamma} (> E_o,t) = \frac{3 q_{\gamma} B^2 \xi {(\kappa C_1)}^{5-2s} m^3 }{2 (3-s)(5m-ms-2)\beta \mu m_p d^2} t^{5m-2ms-2}
\label{eq:hadron}
\ee

\noi where $\mu$ is the mean molecular weight and $m_p$ is the proton
mass.  This formula relates the hadronic emission from the SNR to the
SNR properties. Since we do not generally expect that young SNRs will
be interacting with dense clouds, the target mass used here refers
only to the mass of the material swept-up by the SNR shock wave.

\subsection{Core-Collapse SNRs}
\label{sec:cc}

Core-collapse SNe arise from the explosion of massive stars, with
initial mass generally greater than 8 $\msun$. These stars lose a
considerable amount of mass, modifying the medium around them, and
forming wind-blown bubbles \citep{Weaver1977}. The structure of these
bubbles, and the evolution of SNe within them, has been discussed by
many authors \citep{Chevalier1989, tenorioetal1990, tenorioetal1991,
  Dwarkadas2005, Dwarkadas2007c, vvd08}. The basic structure of a
wind-blown bubble, going outwards in radius from the star, consists of
a freely expanding wind region ending in a wind-termination shock, a
shocked wind region, a contact discontinuity separating the shocked
wind from the shocked ambient medium, an outer shock and the unshocked
ambient medium \citep{vanMarle2006, ta11, dr13}. The crucial point
here is that close in to the star, the SNR will expand in a wind
region. In the simplest approximation, if the wind parameters are
constant, the wind density will decrease outwards in radius as
r$^{-2}$, i.e. with s=2. The parameter $B = \dot{M}/(4 \pi v_w)$ where
$\dot{M}$ is the wind mass-loss rate and $v_w$ is the wind
velocity. The values of $\dot{M}$ and $v_w$ can vary widely depending
on the progenitor star, and the phase of evolution.

If we put $s=2$ in equation \ref{eq:hadron} we get, for a SNR evolving
in a wind medium:

\begin{eqnarray}
F_{\gamma} (> E_o,t) & = & \frac{3 q_{\gamma} B^2 \xi {(\kappa C_1)} m^3 }{2(5m-ms-2)\beta \mu m_p d^2} t^{m-2} \\
& = & \frac{3 q_{\gamma} \xi {(\kappa C_1)} m^3 }{32 {\pi}^2 (3m-2)\beta \mu m_p d^2}\left[\frac{\dot{M}}{v_w}\right]^2 t^{m-2}
\label{eq:pionwind}
\end{eqnarray}

Thus for a SNR evolving in a wind medium, the hadronic emission is
always decreasing in time (after an initial period where the maximum
energy is reached \citep[see for e.g.][]{vt09, dtp12b}).  The reason
for this can be identified from the fact that the swept-up mass is
increasing only as a factor of $R \propto t^m$, whereas the energy
density of cosmic rays is here taken to be decreasing as $t^{-2}$
(assuming $\xi$ to be a constant).

In general the ejecta profile $n$ is expected to vary between 9 and 12
\citep{matzner99}. Thus the expansion parameter $m$ will lie between
0.85 and 0.9, and therefore the emission decreases at a rate between
$t^{-1.15}$ and $t^{-1.1}$.

As can be seen from equation \ref{eq:pionwind} the emission due to
pion decay is a function of the wind mass-loss rate divided by the
wind velocity. This quantity can vary by over two orders of magnitude
depending on the SNR progenitor star, and thus at the same age the
hadronic emission can vary by up to 4 orders of magnitude depending on
the progenitor star. It will be highest for stars that have a high
mass-loss rate and low wind velocity. Stars that fall into this
category would be Red Supergiant (RSG) stars, the progenitors of Type
IIP, and perhaps IIb, SNe. Since they can have mass-loss rates as high
as 10$^{-4} \msun {\rm yr}^{-1}$, and wind velocities of order 10 km
s$^{-1}$, the value of $\dot{M}/{v_w} = 6.35 \times 10^{15}$. On the
other hand, Wolf-Rayet (W-R) stars can have slightly lower mass-loss
rates of about 10$^{-5} \msun {\rm yr}^{-1}$ and wind velocities of
order 2000 km s$^{-1}$, leading to $\dot{M}/{v_w} = 3.17 \times
10^{12}$. Although there will be some differences in the other
parameters, this potentially could lead to more than 5 orders of
magnitude difference in the emission. Thus we would expect that Type
IIP SNe, which arise from RSG progenitors, would have the largest
hadronic luminosity. The caveat though is that since RSG winds have a
small velocity, the RSG wind region cannot extend far out from the
star. For a RSG stage that last about 2 $\times 10^5$ years, and a
wind velocity of 10 km s$^{-1}$, the region would extend only just
over 2 pc.

The results indicate that, as the emission is decreasing with time,
the best time to observe young core-collapse SNe evolving in winds
would be early on. Very early observations in the TeV range can be
attenuated by the pair production process \citep{vt09}, but according
to that paper it would only be important in the first year or so.

\subsection{Type Ia SNRs}

Many of the better known young SNRs that have been observed in
gamma-rays appear to be of Type Ia. These include Tycho, SN 1006,
Kepler and RCW 86. Type Ia SNRs arise from the thermal deflagration of
white dwarfs, and are not expected to considerably modify their medium
(although see \citet{williamsetal11} for the case of SNR RCW 86). Thus
as a first approximation we may assume that they evolve in a constant
density medium, and $s=0$.

Unfortunately, although the power-law density profile assumed for
core-collapse SNe has also been used to describe Type Ia's, it is not
a good representation. \citet{dc98} showed that an exponential density
profile much better represents profiles of Type Ia SNe calculated from
Type Ia explosion models.  A power-law solution with $n=7$, which has
often been used, was shown by \citet{dc98} to not correctly represent
the density and temperature distribution of the material in the
shocked region. Use of an exponential adds one more variable, and
therefore the resultant solution is not self-similar, and not
analytically tractable. Thus it is not possible to employ an
exponential profile and derive a solution analogous to the one above.

As shown in \citet{dc98}, while there is some difference in the radius
and velocity of the outer shock calculated with a power-law as
compared to an exponential density ejecta, the difference is much
larger for the reverse shock parameters. If we consider only the outer
shock and wish to calculate the radius, energy and mass of shocked
material as we need here, the power-law profile may be used to give an
answer possibly correct to order of magnitude or somewhat better,
keeping the above caveats in mind.

If we set $s=0 $ in equation \ref{eq:hadron}, we get that:

\be
F_{\gamma} (> E_o,t) = \frac{3 q_{\gamma} \xi {(\kappa C_1)^5} m^3 }{6 (5m-2)\beta \mu m_p d^2} \rho_{am}^2 t^{5m-2}
\label{eq:pionconst}
\ee

Note that for a constant density medium the hadronic emission is
increasing with time in the ejecta-dominated stage. In fact it has the
same time dependence as the available energy in our formulation. This
is not hard to understand, given that both the swept-up mass as well
as the volume are increasing by $R^3$, (and we have assumed that $\xi$
is constant), and therefore their time-dependence cancels out. Thus
the hadronic emission increases mainly because the amount of energy
available at the shock front that can be used to accelerate protons
increases with time. We would expect that young SNRs which are
expanding in a constant density medium would show a continual increase
in the hadronic emission. Note that qualitatively this statement
should hold true notwithstanding the ejecta profile that is used.

\subsection{SNRs in Wind Bubbles}

As mentioned above, the surroundings of SNRs are much more
complicated. In general core-collapse SNRs would first expand in a
wind medium, then encounter a wind termination shock followed by a
more or less constant density medium. In the case of a Type Ic SN the
wind density at the termination shock would likely be a factor of 4
lower than that of the shocked constant density wind medium; thus the
SNR is subsequently evolving into a higher density medium. In the case
of a IIP/IIb SN, the density in the wind could be much higher than
that in the constant density medium, and the SN is evolving from a
high density to a lower density medium.

It would be tempting to assume that while the SNR is expanding in
either the freely expanding wind medium or the shocked wind medium,
the above expressions could be used to describe it. The difficulty is
that for a time which can be several doubling times of the radius, as
the SN shock crosses the wind termination shock or a discontinuity,
the SNR is in a transition stage, and no longer able to be described
by the self-similar solutions given above. This is much more
accentuated in the case where the jump in density from the wind to the
main-sequence shocked bubble is large. This can happen in Type IIP
SNe, when crossing from the RSG wind to the main-sequence bubble. The
shocked structure then differs from the self-similar solution
\citep[see][]{tdp13a}, limiting the applicability of the above
formulae. However, we would expect that although the actual radius
evolution with time cannot be easily described, the general trend
regarding the increase or decrease of hadronic emission would still
hold.

In cases where the density jump, and transition time, may be smaller,
say Type Ic SNe, the solution may reach the self-similar structure
quickly after the transition, and the above results may be more easily
attained. Therefore we would expect the hadronic emission to first
decrease with time as the SNR evolves in the wind region, then undergo
a period of transition, and then gradually start to increase in
time. This trend is also seen in the numerical calculations of
\citet{caprioli11}. If the SNR shock goes on to collide with the dense
shell surrounding the wind-bubble, or comes close enough to it, the
accelerated particles may impact the dense shell, giving a further
increase in the hadronic emission.

\subsection{SNRs in Time-Dependent Winds}

In \S \ref{sec:cc} we assume that the wind which the SNR was evolving
in had constant wind parameters, namely the mass-loss rate and wind
velocity. There is of course no good reason why the wind parameters
should be constant. As stars evolve on the main-sequence and beyond,
their mass-loss properties change with time. Especially towards the
end of their lifetime, the surroundings of many observed SNe have
indicated the presence of time varying winds \citep{mccray1993,
  chugaietal04, mccray2007, smithetal08, ddb10}.  The X-ray light
curves of young SNe suggest that many of them evolve in winds that do
not appear to have constant wind parameters \citep{dg12}. The further
out in radius one goes, the more likely it is that the wind parameters
may not be constant. In practice this means that the circumstellar
density will not decrease as r$^{-2}$.

In principle, if the winds are time-dependent, then we have to go back
to equation \ref{eq:pion} and insert the time-dependent part into the
integral and re-calculate. To estimate the effect of time-dependent
winds, one could assume that the time dependence for instance somehow
conspired to give a value of $s$ intermediate between the wind and
constant density case. For a sample case $s=1$, we can write:

\be F_{\gamma} (> E_o,t) = \frac{3 q_{\gamma} B^2 \xi {(\kappa
    C_1)}^{3} m^3 }{4(4m-2)\beta \mu m_p d^2} t^{3m-2}
\label{eq:pioninter}
\ee

Note that in this case the emission is increasing for $m > 2/3$ and
decreasing for lower values of $m$ as the remnant decelerates. The
important point is that it may no longer be monotonically increasing
or decreasing throughout the ejecta-dominated stage, but reaches a
maximum and then begins to decrease.

\subsection{Reverse Shocked Emission:} 

The above analysis mainly considered particles accelerated at the
forward shock. In young SNRs that have not reached the Sedov or
adiabatic stage, a reverse shock exists that expands back into the
ejecta. Whether particles can be accelerated at the reverse shock or
not is an ongoing question. There have been several suggestions of
non-thermal emission arising from the reverse shock \citep{dkrd02,
  rdbr02}, the most convincing one being for the case of Cas A
\citep{hv08}. Since we know that the reverse shock is also a
collisionless shock and must be mediated by magnetic fields, one
cannot deny the existence of a magnetic field, however small, at the
reverse shock. The question then is whether the field can be amplified
enough to accelerate particles. In a series of papers, Telezhinsky et
al.~\citep{tdp12a,tdp12b,tdp13a} have computed the acceleration of
particles in SNRs, assuming acceleration at both forward and reverse
shock. They show that due to the larger number of particles but lower
maximum energy at the reverse shock, the resultant integrated spectrum
from the SNR will be steeper (softer) at TeV energies than at GeV
energies, which is consistent with the observations.

If particles are accelerated at the reverse shock, it may be
especially important in situations where the reverse and forward
shocks are expanding in regions of different densities. This may occur
for instance when the SNR crosses a discontinuity or wind termination
shock, such as in wind bubbles. The reverse shock may be in a region
with different density compared to the forward shock, and may even
dominate the emission.

In the above calculations it is possible to take the reverse shocked
emission into effect, or to assume emission from both forward and
reverse shocked plasma. The calculations are simplified in the thin
shell approximation, where the region between the forward and reverse
shocks is assumed to be small compared to the radius of the
remnant. This requires a few changes in equation \ref{eq:pion}. For
$\kappa$ we substitute ${\kappa}_r < 1 $ for the reverse shock. In the
thin shell approximation, the mass swept up by the reverse shock can
be written in terms of that swept-up by the forward shock as
\citep{nfk06}:

\be
M_{rev} = \frac{n-4}{4-s} M_{cs}
\ee

\noi
Similarly, the density behind the reverse shock can be written in
terms of the density behind the circumstellar shock:

\be 
\rho_{rev} = \frac{(n-3)(n-4)}{(3-s)(4-s)}\,{\rho}_{cs} 
\ee

With the appropriate changes, we get the hadronic emission from the
reverse shock to be:

\be {F_{\gamma}}_R (> E_o,t) =
\frac{(n-4)^2(n-3)}{(4-s)^2(3-s)^2}\frac{3 q_{\gamma} B^2 \xi
  {({\kappa}_r C_1)}^{5-2s} (1-m)^3 }{2(5m-ms-2)\beta \mu m_p d^2}
t^{5m-2ms-2}
\label{eq:hadronrev}
\ee

If the thin-shell solution is not applicable, then one needs to go
back to equation \ref{eq:pion} and work through, while inputting the
actual reverse shock parameters.

\section{Application to Individual SNRs}

The validity of these solutions can be gauged by computing the
emission for observed SNRs. Unfortunately, there are not many young
core-collapse SNRs that have been observed at GeV and TeV energies. As
pointed out above, many of the observable ones appear to be of Type
Ia, although SN surveys suggest that core-collapse ones should be more
common. This could be interpreted in the context of our above results
as saying that since the emission from core-collapse SNe decreases
with time, one would be less likely to see them after several hundreds
of years, unless they are expanding in a dense wind.

\subsection{Cas A} 
One of the SNRs that is assumed to be expanding in a dense wind, and
has been detected at high energies, is Cas A. \citet{co03} suggest
that the SNR is expanding in a RSG wind, with a wind mass-loss rate of
2 $\times 10^{-5} \msun {\rm yr}^{-1}$ for a wind velocity of 10 km
s$^{-1}$. They infer the energy of the explosion to be about 4 times
the standard energy ($4 \times 10^{51}$ ergs), and the ejecta to have
a density profile that goes as r$^{-10.12}$, in accordance with the
work of \citet{matzner99} for a massive star with a radiative
envelope. \footnote{In actual fact, \citet{co03} use the density
  profile for a red supergiant star with a radiative envelope given in
  \citet{matzner99}, which is somewhat different from that assumed
  here. That profile has a density power-law density that starts off
  as steep and slowly becomes less steep. This will cause some
  difference in the results} This gives $m \sim 0.88$. Using these
parameters (with $n=10$), we can calculate the value of $C_1$ to be
1.165 $\times 10^{10}$, and take $\kappa \sim 1.2$, and $\beta =
0.5$. We assume the fraction of energy transferred to cosmic rays to
be 0.1. The value of $q_{\gamma} > 100 $ MeV is obtained from DAV94 to
be 0.5 $\times 10^{-13}$, almost independent of the spectral index
$\alpha$ at low energies. We take $\mu=1.4$. Using these values we get
from equation \ref{eq:pionwind} that

\be
{F_{\gamma}}_{CAS A} (> 100\;{\rm MeV}) = 5.5 \times 10^{-8} \; {\rm cm}^{-2}\, {\rm s}^{-1}
\ee

This is about 6 times larger than the flux reported by {\it Fermi}
\citep{fermicasa}, which is the flux above 500 MeV, and so may be
expected to be slightly lower. However, the {\it Fermi} best fit
suggests that at most 2\% of the total energy has gone into cosmic
rays. If we assume $\xi=0.02$ then we get a decrease of a factor of 5,
giving a flux ${F_{\gamma}}_{CAS A} (> 100 MeV) = 1.1 \times 10^{-8} $
that is comparable to the {\it Fermi} result. Furthermore, the {\it
  Fermi} result assumes a total energy half that of \citet{co03},
which would reduce the flux further, although not exactly by a factor
of 2 since other parameters would also change. However this exercise
suggests that within the error bars, the flux computed via this method
is close to that detected by {\it Fermi}, and therefore suggests that
the $\gamma$-ray emission from Cas A could be due to hadronic
processes. A similar inference was made in \citet{fermicasa}.

\subsection{SN 1993J} 
SN 1993J is one of the closest SNe to have exploded in the past couple
of decades, and one of the most observed in X-ray and radio. It is not
surprising that several papers have attempted to model the
$\gamma$-ray emission from SN 1993J \citep{kdb95, vt09}.  \citet{vt09}
made a very detailed computation of the particle acceleration, radio
emission and hadronic emission from SN 1993J. He finds (eqn.~53) that
the emission goes as $t^{-1}$. This is primarily because he assumes
that $\xi_{CR}$, the ratio of the cosmic-ray pressure to gas pressure,
goes as $V_{sh}^{-1} \propto t^{1-m}$, and therefore the cosmic ray
energy density decreases as $t^{-(m+1)}$. As shown above, in the case
of core collapse SNe, we have assumed that the cosmic ray energy
density decreases as $t^{-2}$, the same as the internal energy density
and therefore the gas pressure. Thus the difference between the time
evolution in the two cases is due to the fraction of energy going into
cosmic rays, and its constancy with time.

We can compute the $\gamma$-ray flux from SN 1993J, using the
parameters assumed by \citet{vt09}, namely m=0.83, B=1.9 $\times
10^{14}$ g cm$^{-1}$, d=3.63 Mpc, and $\kappa C_1 = 2.79 \times
10^{10}$. We then get from equation \ref{eq:pionwind}, assuming
$\mu=1.4$, that the flux from SN 1993J in the TeV range goes as

\be {F_{\gamma}}_{93J} (> 1\; {\rm TeV}) = 2.14 \times 10^{-11} \rm
t_{days}^{-1.17}\; {\rm cm}^{-2}\, {\rm s}^{-1} 
\ee

\noindent
The constant in front is about an order of magnitude larger than in
\citet{vt09}, but the flux is decreasing somewhat faster, as
t$^{-1.17}$ rather than t$^{-1}$.

The flux greater than 100 MeV is then:

\be {F_{\gamma}}_{93J} (> 100\; {\rm MeV}) = 1.07 \times 10^{-7} \rm
t_{days}^{-1.17}\; {\rm cm}^{-2}\, {\rm s}^{-1} 
\ee

After 20 years the flux in the {\it Fermi} range is 3.23 $\times
10^{-12} {\rm cm}^{-2}\, {\rm s}^{-1}$, which is below the level that
can be detected by {\it Fermi}.

It is not clear however that the density of the medium around SN 1993J
does decrease as r$^{-2}$, and that a constant mass-loss rate is
appropriate. Early reports \citep{vandyketal94} inferred a density
decline of r$^{-1.5}$ from the radio emission, and r$^{-1.7}$ from the
X-ray emission \citep{sn95}. A circumstellar medium density decreasing
as r$^{-1.5}$ was used by \citet{kdb95} to compute the emission from
SN 1993J. Inspired by these results, \citet{mdb01} modelled the
hydrodynamic interaction, and found that the radio emission was better
explained by a complicated density profile that diverged from
r$^{-2}$. On the other hand, detailed calculations of the radio
emission, encompassing much more physics, were carried out by
\citet{fb05}, who suggested that the radio could be explained by the
self-similar solution embodying shock expansion into a steady wind, a
result that was echoed by \citet{vt09}. However, \citet{nymarketal09}
then found that the self-similar solution was not adequate to explain
the X-ray emission, and used the ejecta profile in
\citet{sn95}. Interpretation of the data does not lead to conclusive
results \citep{bietenholzetal10}, further complicating the situation.

In this context it is interesting that the hadronic emission, and its
time evolution, differ considerably depending on whether the medium
goes as r$^{-2}$ or r$^{-1.5}$. One may speculate that in future, if
nearby SNe could be detected and followed with advanced instruments at
very high energies, they may be combined with other multi-wavelength
results to help infer the nature of the medium that SNRs are expanding
in, and thereby the nature of mass-loss from the progenitor star.

\subsection{SN 1987A}

Even though SN 1993J happens to be one of the closer core-collapse SNe
of the modern era, the large distance to the SN renders the flux small
enough to be practically undetectable with current telescopes. It is
therefore opportune to turn our attention to the closest SN in modern
times, SN 1987A, which exploded in the LMC at a distance of about 50
kpc. The ambient medium into which the SN shock is expanding is
distinctly non-uniform, as revealed by X-ray, optical, and radio
observations, and quite unlike other core-collapse SNe that we have
discussed. In order to explain the increasing radio and X-ray
emission, \citet{cd95} suggested that, after about 3 years of
expansion in a wind medium, the SN shock impacted a dense HII region
formed by the progenitor winds and ionized by the pre-SN star. Finally
(after about 25 years or so), the SN shock will begin to interact with
the dense equatorial ring of material that is seen so beautifully in
optical Hubble Space Telescope images.  The density profile into which
the shock is expanding is thus quite inhomogeneous. However, and since
we are mainly interested in the mass of material that has been
shocked, we can obtain a rough estimate of the flux by neglecting the
mass of the swept-up wind, and assuming that up to about 26 years the
SN shock is interacting with a constant density HII region with mean
hydrogen density of about 200, which gives approximately the correct
mass. We use equation~\ref{eq:pionconst} to compute the flux from the
SN, assuming the parameter $n=9$, and ejecta parameters as used in
\citet{deweyetal12}. The flux greater than 100 MeV, at an age of 26
years, is then:

\be {F_{\gamma}}_{87A} (> 100\; {\rm MeV}) = 4.04 \times 10^{-10} {\rm
  cm}^{-2}\, {\rm s}^{-1} 
\ee

The observed synchrotron power-law index of SN 1987A, as measured in
the radio, is very soft \citep{zanardoetal13}, and varies over the
remnant, with values 2.4 to 2.8. At TeV energies, as shown by
\citet{dav94}, the emissivity $q_{\gamma}$ can vary by over two orders
of magnitude as the spectrum steepens from a power-law index of 2.1 to
2.6. However, at the shock front, within the assumption of Bohm
diffusion, we would expect the spectrum at very high energies to be
close to 2. In fact non-linear DSA predicts a harder spectrum. The
{\em observed} spectrum may be softened due to other effects such as
Alfvenic drift \citep{zp08,caprioli11}. Therefore, we assume a
spectral index at the shock close to 2, and a gamma-ray emissivity of
$q_{\gamma} \approx 1 \times 10^{-17}$, which gives:

\be 
{F_{\gamma}}_{87A} (> 1\; {\rm TeV}) = 8.1 \times 10^{-14}  {\rm cm}^{-2}\, {\rm s}^{-1} 
\ee

This is a few times smaller than that calculated by
\citet{bkv11}. This is to be expected given the difference in their
assumptions about the surrounding medium, and the nature of our
estimate.  In fact it is reassuring that they are comparable.  One of
the problems with our approach is in fact that since the SN is first
evolving in a wind and then in a circumstellar medium, as pointed out
earlier, the shock structure is going to be in a transition state for
the first few years after impact with the HII region
\citep{Dwarkadas2007b, deweyetal12}. Nevertheless, this does provide
us with an order of magnitude estimate of the hadronic flux. Note that
in our case the flux is increasing with time as t$^{1.33}$.

The hadronic flux level of SN 1987A is at present
undetectable. However, it is exciting to note that the SN will soon,
if it has not already, start interacting with the dense equatorial
ring formed by the progenitor star, with density n$_H \sim 10^4$
\citep{lundqvist1999}. Then, depending on the thickness of the ring,
the $\gamma$-ray flux is expected to increase significantly over the
next decade. We can approximate the luminosity at 36 years assuming an
average density of about n$_H \sim 2 \times 10^3$ into which the SN
shock has been expanding, again keeping in mind that the self-similar
solution may not be entirely appropriate. We then get for the flux at
36 years to be:

\be {F_{\gamma}}_{87A} (> 100\; {\rm MeV}) = 1.9 \times 10^{-8} {\rm
  cm}^{-2}\, {\rm s}^{-1} \ee

and 

\be {F_{\gamma}}_{87A} (> 1\; {\rm TeV}) = 3.8 \times 10^{-12} {\rm
  cm}^{-2}\, {\rm s}^{-1} \ee

The latter is detectable with an instrument such as the current HESS
telescope array in less than an hour. Even if the number is off by a
factor of 5, which is quite possible given the inhomogeneous and
complex nature of the surrounding medium, it should still be
detectable within a few hours. In fact it is possible that a 25 hour
observation could detect it a few years earlier. All in all, even
these order-of-magnitude estimates suggest that SN 1987A should be
kept in mind as a potential gamma-ray source in the next decade, whose
detection would strongly support the model for cosmic-ray acceleration
in SNRs.

SN 1987A is unusual in that it is a core-collapse SN whose hadronic
luminosity is expected to increase with time early on, whereas for
most SNe it is decreasing with time. This had previously been pointed
out, albeit with different time dependence due to different
assumptions of velocity and density, by \citet{kdb95}. The increasing
flux in the radio and X-ray regimes has been monitored for the past
two decades, and detection at very high Gev-TeV energies would serve
to corroborate our understanding of both the hadronic emission and the
details of the SN environment. Unfortunately, the {\it Fermi} space
telescope will probably not be active by the time the SN brightens
sufficiently, but ground based ACTs such as HESS should be able to
detect it in the TeV range. It also promises to be an exciting target
for the upcoming Cherenkov Telescope Array, which, if our estimates
are good to even an order of magnitude, should definitely detect it.

\section{Discussion:} 

The above results indicate that if we take a statistical ensemble of
young core-collapse SNRs with age, one would see a gradual decrease in
the hadronic emission with age, varying between $t^{-1.1}$ and
$t^{-1.15}$, but with a scatter that extended to about 2 orders of
magnitude in each direction. IIp's arise from progenitor stars that
are much lower in mass than 1b/c's, and are larger in number, given
the weighting of the initial mass function. In theory they would be
expected to dominate, thus leading to a high luminosity that decreases
more or less as t$^{-1.1}$. Unfortunately, because of the very few
young SNRs that are observed in our galaxy, the small-number
statistics do not present a true picture. Furthermore, even among
those that are seen, many of them are Type Ia, which should show
hadronic emission that increases with time. We would expect therefore
that a plot of luminosity v/s age for SNRs would initially decrease
(due to the predominance of core-collapse SNe) but then begin to
increase as the Type Ia's become brighter, and the core-collapse ones
begin to expand in a constant density medium. However, unless the
statistics can be improved by future telescopes, perhaps detecting
young SNRs such as SN 1987A in galaxies outside our own, it will be
hard to get a good statistical description.

\citet{vt09} carried out a highly detailed exploration of the particle
acceleration in SN 1993J. The differences between our simpler estimate
and his detailed computation can be understood in terms of the
differences in the evolution of the cosmic ray pressure, and suggest
that our estimates are reasonable in calculating the luminosity or
flux good to a factor of a few.

\citet{kdb95} calculated the evolution of the hadronic photon
luminosity for two SNe. Their results can easily be understood in
terms of the expressions derived herein. They assumed that in the
ejecta-dominated phase, the shock velocity is a constant ($m=1$). That
implies from equation \ref{eq:hadron} that the flux will decrease as
$F_{\gamma} \propto t^{5-2s-2}$. For a SN expanding in a wind, as they
assumed for SN 1987A, with $s=2$ we then get that the flux decreases
as t$^{-1}$, as they found. In the case of SN 1993J, they assumed it
to be expanding in a medium whose density goes as $\rho \propto
r^{-1.5}$, i.e. $s=3/2$, in which case, with $m=1$, we get that
$F_{\gamma} \propto t^{0}$, i.e. the flux is a constant with
time. Thus our results reduce to theirs in the appropriate parameter
range, further emphasizing their applicability and versatility.

In this context it must be mentioned that by taking the quantity $\xi$
out of the integral, we are in effect replacing the instantaneous
energy loss to cosmic rays with the total energy lost over a finite
period of time, thus making $\xi$ a global efficiency parameter, and a
proxy for the total energy lost over a sufficiently large period such
as the ejecta-dominated phase. It seems even more plausible then to
take $\xi=0.1$, given that about 10\% of the total SN kinetic energy
is needed to account for the total flux of cosmic-rays up to the knee
of the spectrum.

\subsection{Energy in Cosmic Rays} 
A question that naturally arises is whether the self-similar solution
used herein would be modified due to the particle acceleration. Many
authors have shown that the shocked region will contract further as
more and more kinetic energy is diverted to cosmic-ray acceleration
\citep{be01, epsbg07, fds12}, and the density jump across the shock
will change.  However, as long as the ratio of cosmic-ray to gas
pressure does not exceed 10\%, the test particle solutions, and the
unmodified shock structure, should be reasonably correct
\citep{kang10}. This was addressed by \citet{chevalier83}, who showed
that while there will be some effect on the radii and densities, it
will be only at a few percent level if 10\% of the kinetic energy goes
into accelerated particles. (As an aside, we note that although the
energy expended in cosmic ray acceleration is calculated differently
in \citet{chevalier83}, the resulting time-dependence, although
expressed in terms of $n$ and $s$ rather than $m$, is exactly the same
as reported in this paper.) Thus, using $\xi=0.1$, it is reasonable to
assume that the shock structure is reasonably well represented by the
self-similar solutions.

The question that remains then is whether the cosmic-ray pressure
exceeds the gas pressure by 10\%, or equivalently whether a large
amount of energy is expended in accelerating cosmic
rays. Unfortunately there is no easy answer to that. A simple
calculation \citep[for e.g.][]{longairv3} shows that about 10\% of the
total energy in SNRs is required to fuel the galactic cosmic-ray
flux. Therefore it does not seem that a much higher amount is
necessary to explain the observed cosmic-ray flux. Analysis of SNR
evolution, including particle acceleration, provide conflicting
numbers. For Cas A, \citet{pf09} suggest that about 30\% of the energy
goes into accelerating cosmic rays, while the broadband fits by the
{\it Fermi} team find that the energy content in cosmic rays is less
than 2\% of the total explosion energy. In the case of Tycho's SNR,
\citet{kbv11} find that about 10-20\% of the energy is lost in
cosmic-ray escape; \citet{mc12} suggest that 5-10\% of the explosion
energy goes into cosmic rays; \citet{zclz13} suggest that the energy
conversion efficiency is only 1\% but that molecular-cloud interaction
is involved; \citet{bkv13} require a two-phase inhomogeneous medium;
while \citet{ad12} require a very small efficiency but show that using
a multi-zone model, leptonic emission remains a possibility to explain
the very high energy emission. Therefore, in the case of Tycho's SNR,
no consensus exists on even the emission mechanism, let alone the
energy needed. In the case of SN 1006, \citet{bkv12} find that the
energy in accelerated cosmic rays is about 5\% of the total explosion
energy, although this could double over time, while \citet{kbv11} find
that energy losses due to cosmic ray escape are 20-50\% of the kinetic
energy flux.

Studies of other SNRs are equally conflicting. \citet{vt09} found that
about 19\% of the energy processed by the shock goes into cosmic-ray
energy over the first 8.5 years, but yet the shock is only weakly
modified. This can perhaps be understood by the fact that the energy
processed by the shock is only a small percentage of the total kinetic
energy available, so the fraction of the cosmic-ray energy to total
energy is much smaller. Similarly, \citet{epsr10} find that the
instantaneous energy that goes into cosmic-rays can be as high as
25-50\% for SNR RX J1713, but that the overall energy is about 13\%
for the leptonic model that they favor.

Another observation supposedly pointing to efficient cosmic ray
acceleration is the ratio of the forward shock (blast wave) to the
contact discontinuity. For e.g. in Tycho's SNR, \citet{warrenetal05}
find that the ratio of the blast wave to the contact discontinuity is
0.93, which is much larger than that expected from adiabatic
hydrodynamic models, and suggests that the blast wave is much closer
to the contact discontinuity than expected. They take that as a sign
that the region has been compressed due to efficient cosmic-ray
acceleration. However, this presumes that the shock wave has evolved
over a constant density medium throughout its lifetime, which is not
certain. Furthermore, \citet{orlandoetal12} clearly show that the
smaller separation can be easily reproduced by clumping of the ejecta,
and does not require cosmic-ray acceleration. It is significant that
the ratio of forward to reverse shock can be reproduced by
hydrodynamical models, it is only the radius of the contact
discontinuity that seems contrary to expectations, thus further
supporting the claim of \citet{orlandoetal12}.

Do the non-linear DSA models produce spectra that fit the observations
better? Since the low-energy particles see a smaller shock jump, and
the higher energy particles a larger shock jump, the models actually
produce spectra that are steeper at GeV than at TeV energies, which is
the inverse of the observations, which are far steeper at TeV
energies. However, as shown by \citet{caprioli11}, this can be
modified to match the observations by assuming that the Alfvenic drift
of magnetic scattering centers plays a role.

On the other hand, as shown by \citet{tdp12a, tdp12b, tdp13a}, softer
spectra in the TeV regime that match the observations can be obtained
even in test-particle mode, by making the simple assumptions that both
forward and reverse shocks can accelerate particles. The higher
intensity and lower maximum energy of particles accelerated at the
reverse shock, when combined with the forward shock accelerated
particles, produce integrated particle spectra that are softer at TeV
wavelengths than in the GeV region, consistent with the observations.

It seems, at least at present, that there is neither motivation nor
compulsion for a large amount of explosion energy ($ > 10\% $) to be
diverted to cosmic-ray acceleration, nor is there any specific clear
observation of the same.

\subsection{Circumstellar Environments} 
As shown above, the hadronic emission goes as $\rho^2$ for young
SNRs. \citet{dav94} have an expression for hadronic emission with one
power of density in it. The second arises from the fact that the
energy processed instantaneously at the shock is much smaller than the
total SNR energy. This dependence on the square of the density is
similar to that for thermal X-ray emission from SN shock waves
\citep[see for e.g.][]{dg12}, therefore it is no surprise that
hadronic emission is pre-dominant in SNRs which also show a high
intensity of thermal x-ray emission.

\citet{caprioli11} had modelled the hadronic emission from SNRs
interacting with ``non-homogeneous circumstellar environments'',
basically a wind-blown bubble. That model though has many problems, as
is evident by comparing to analytical and numerical models of
wind-bubbles \citep{glm96, Dwarkadas2005, Dwarkadas2007c, ta11,
  vanMarle2006, dr13}: (1) Firstly, it assumes a RSG wind close in to
the star with a density profile that goes as r$^{-2}$, followed by a
W-R bubble with constant density. As shown by many authors, the RSG
wind also blows a shell, but due to the much larger velocity and
momentum of the W-R wind, the latter completely destroys the RSG shell
and pushes the RSG material further out, mixing it up in the
process. The RSG and W-R portions do not exist separately as used in
\citet{caprioli11}. In particular, there is no RSG wind at the end of
the W-R stage, but a W-R wind, whose density would be orders of
magnitude lower. (2) A RSG wind going directly into a constant density
W-R medium without an intervening discontinuity such as a shock is in
fact not even a hydrodynamically stable situation, and cannot exist in
practice. (3) \citet{caprioli11} has a high-density jump at the end of
the bubble, after which the density stays constant. In practice W-R
bubble shells are bounded by radiative shocks, where the shock jump
can be extremely high. Thus there is a dense shell surrounding the
bubble, following which the ISM density should be much lower than the
shell. The SNR shock may be considerably decelerated at the dense
shell. None of these factors are taken into account. (4) The region is
arbitrarily divided into ejecta-dominated and Sedov-Taylor.  As
mentioned above it takes several swept-up masses before the
Sedov-Taylor stage is reached. In many cases, the Sedov-Taylor stage
will not be reached for evolution in a W-R bubble. It is also very
unlikely that the Sedov stage will be reached by a SN expanding in a
wind medium, given the decreasing density and the fact that a
substantially large mass is needed.  (5) The solution given for the
evolution of the shock wave in the wind medium by \citet{caprioli11}
(their equations 3.4 and 3.5) are for a specific value of the
parameter ``n'' (as used here), which is not generally applicable over
the entire mass range of progenitors. More specifically, it is not
necessarily applicable to W-R stars \citep{matzner99}, to which it was
applied. (6) Finally, there is a large variety in wind mass-loss rates
and velocities as shown above, up to several orders of magnitude, and
therefore a single model with fixed parameters to fit all progenitor
stars is a gross oversimplification that will not work for any
particular star in practice. We think it is this oversimplification
that leads to the bimodal distribution suggested in $\gamma$-ray
hadronic emission found in \citet{caprioli11}.

To be fair, the description of the emission given in this paper is
simplified. On the other hand, we have a realistic treatment of the
surrounding medium and the SN evolution within this medium, which we
believe is a crucial ingredient and essential to understanding the
hadronic emission. Unless the SNR evolution in the ambient medium is
properly treated, even the most accurate treatment of the particle
acceleration and hadronic emission will likely lead to incorrect
results, because as pointed out the emission depends crucially on the
shock radius, velocity and swept-up mass. Our intention as specified
is not to calculate in detail the emission from any given SNR
(although it appears to do that reasonably well as shown), but to
understand the time-evolution, and the factors it depends on, as well
as the role of the circumstellar medium, and to begin to evaluate the
statistical properties of an ensemble of SNRs.

\section{Summary and Conclusions} 

We have derived a simple expression to compute the hadronic emission
from young SNRs, taking the evolution of the SN ejecta into various
environments into account, and assuming that the target mass for
proton collisions is the material swept-up by the SN shock wave(s). We
calculate its dependence on the evolutionary parameters (ejecta energy
and mass-loss rate), the density structure of the surrounding medium,
and the expansion parameter of the shock wave. Our results show that
for core-collapse SNRs expanding into the winds of their progenitor
stars, the hadronic emission must decrease in time for winds with
constant parameters. On the other hand, for expansion into a medium
with a constant density, the hadronic emission is expected to increase
with time. In the appropriate parameter regime, these results can
reproduce the time evolution of the hadronic luminosity calculated for
many specific SNe in the literature, thus demonstrating their broad
applicability.

We have not included the so-called ``interacting'' SNRs, which are
interacting with dense molecular clouds. Most of those are in a much
more evolved stage, and not in the ejecta-dominated stage. In
principle, if they were to be in the ejecta-dominated stage, it is not
difficult to include them. The main difference between the
non-interacting remnants discussed here, and the interacting remnants,
is the mass of the material with which they are interacting, and its
distance from the SNR shock wave. That means that we could start from
equation \ref{eq:pion} and, instead of the swept-up mass, use the mass
of the cloud or interacting material and re-derive the equations. A
major difference is that, depending on the distance to the cloud, only
the highest energy cosmic rays would be able to diffuse towards it
\citep{gabici11, tdp12b}.

The equations derived here can provide observers with a simple method
to compute the hadronic emission from any given young SN or SNR, and
thus get a better-than-order-of-magnitude estimate of the hadronic
flux with knowledge of a few basic explosion parameters. Conversely,
knowing that the emission is hadronic, they could be used to derive
the density of the ambient medium or the SN parameters.

In future papers we plan to derive similar expressions for the
leptonic emission. Our goal is to have a suite of formulae that will
provide an aid to understanding the emission from young SNRs in the
age of {\it Fermi} and ground-based IACT's, allow us to estimate the
dependence on various parameters, and therefore help to interpret the
upcoming {\it Fermi} SNR catalog.

\section*{Acknowledgments}

VVD's research on the very high energy emission from young SNRs is
supported by NASA Fermi grant NNX12A057G. Some partial support was
also received from NSF PHYS award 0969529. 

\bibliographystyle{mn2e}
\bibliography{paper}


\bsp

\label{lastpage}

\end{document}

%% file: macros.tex
\newcommand{\vper}{\mbox{${v_{\perp}}$}}
\newcommand{\vpar}{\mbox{${v_{\parallel}}$}}
\newcommand{\uper}{\mbox{${u_{\perp}}$}}
\newcommand{\vperout}{\mbox{${{v_{\perp}}_{o}}$}}
\newcommand{\uperout}{\mbox{${{u_{\perp}}_{o}}$}}
\newcommand{\vperin}{\mbox{${{v_{\perp}}_{i}}$}}
\newcommand{\uperin}{\mbox{${{u_{\perp}}_{i}}$}}
\newcommand{\upar}{\mbox{${u_{\parallel}}$}}
\newcommand{\uparout}{\mbox{${{u_{\parallel}}_{o}}$}}
\newcommand{\vparout}{\mbox{${{v_{\parallel}}_{o}}$}}
\newcommand{\uparin}{\mbox{${{u_{\parallel}}_{i}}$}}
\newcommand{\vparin}{\mbox{${{v_{\parallel}}_{i}}$}}
\newcommand{\dout}{\mbox{${\rho}_{o}$}}
\newcommand{\din}{\mbox{${\rho}_{i}$}}
\newcommand{\da}{\mbox{${\rho}_{1}$}}
\newcommand{\mfast}{\mbox{$\dot{M}_{f}$}}
\newcommand{\mslow}{\mbox{$\dot{M}_{a}$}}
\newcommand{\beqn}{\begin{eqnarray}}
\newcommand{\eeqn}{\end{eqnarray}}
\newcommand{\be}{\begin{equation}}
\newcommand{\ee}{\end{equation}}
\newcommand{\noi}{\noindent}
\newcommand{\ftheta}{\mbox{$f(\theta)$}}
\newcommand{\gtheta}{\mbox{$g(\theta)$}}
\newcommand{\ltheta}{\mbox{$L(\theta)$}}
\newcommand{\stheta}{\mbox{$S(\theta)$}}
\newcommand{\utheta}{\mbox{$U(\theta)$}}
\newcommand{\xitheta}{\mbox{$\xi(\theta)$}}
\newcommand{\vs}{\mbox{${v_{s}}$}}
\newcommand{\ro}{\mbox{${R_{0}}$}}
\newcommand{\pa}{\mbox{${P_{1}}$}}
\newcommand{\va}{\mbox{${v_{a}}$}}
\newcommand{\vo}{\mbox{${v_{o}}$}}
\newcommand{\vp}{\mbox{${v_{p}}$}}
\newcommand{\vw}{\mbox{${v_{w}}$}}
\newcommand{\vf}{\mbox{${v_{f}}$}}
\newcommand{\lprime}{\mbox{${L^{\prime}}$}}
\newcommand{\uprime}{\mbox{${U^{\prime}}$}}
\newcommand{\sprime}{\mbox{${S^{\prime}}$}}
\newcommand{\xiprime}{\mbox{${{\xi}^{\prime}}$}}
\newcommand{\mdot}{\mbox{$\dot{M}$}}
\newcommand{\msun}{\mbox{$M_{\odot}$}}
\newcommand{\yr}{\mbox{${\rm yr}^{-1}$}}
\newcommand{\kms}{\mbox{${\rm km} \;{\rm s}^{-1}$}}
\newcommand{\lambdav}{\mbox{${\lambda}_{v}$}}
\newcommand{\lequ}{\mbox{${L_{eq}}$}}
\newcommand{\eqpratio}{\mbox{${R_{eq}/R_{p}}$}}
\newcommand{\ra}{\mbox{${r_{o}}$}}
\newcommand{\bfig}{\begin{figure}[h]}
\newcommand{\efig}{\end{figure}}
\newcommand{\tone}{\mbox{${t_{1}}$}}
\newcommand{\done}{\mbox{${{\rho}_{1}}$}}
\newcommand{\dsn}{\mbox{${\rho}_{SN}$}}
\newcommand{\dzero}{\mbox{${\rho}_{0}$}}
\newcommand{\ve}{\mbox{${v}_{e}$}}
\newcommand{\vej}{\mbox{${v}_{ej}$}}
\newcommand{\Mch}{\mbox{${M}_{ch}$}}
\newcommand{\mej}{\mbox{${M}_{e}$}}
\newcommand{\Mst}{\mbox{${M}_{ST}$}}
\newcommand{\dam}{\mbox{${\rho}_{am}$}}
\newcommand{\Rst}{\mbox{${R}_{ST}$}}
\newcommand{\Vst}{\mbox{${V}_{ST}$}}
\newcommand{\Tst}{\mbox{${T}_{ST}$}}
\newcommand{\no}{\mbox{${n}_{0}$}}
\newcommand{\Efif}{\mbox{${E}_{51}$}}
\newcommand{\rsh}{\mbox{${R}_{sh}$}}
\newcommand{\msh}{\mbox{${M}_{sh}$}}
\newcommand{\vsh}{\mbox{${V}_{sh}$}}
\newcommand{\vrev}{\mbox{${v}_{rev}$}}
\newcommand{\rpr}{\mbox{${R}^{\prime}$}}
\newcommand{\mpr}{\mbox{${M}^{\prime}$}}
\newcommand{\vpr}{\mbox{${V}^{\prime}$}}
\newcommand{\tpr}{\mbox{${t}^{\prime}$}}
\newcommand{\cone}{\mbox{${c}_{1}$}}
\newcommand{\ctwo}{\mbox{${c}_{2}$}}
\newcommand{\cthree}{\mbox{${c}_{3}$}}
\newcommand{\cfour}{\mbox{${c}_{4}$}}
\newcommand{\Te}{\mbox{${T}_{e}$}}
\newcommand{\Ti}{\mbox{${T}_{i}$}}
\newcommand{\Ha}{\mbox{${H}_{\alpha}$}}
\newcommand{\Rprime}{\mbox{${R}^{\prime}$}}
\newcommand{\Vprime}{\mbox{${V}^{\prime}$}}
\newcommand{\Tprime}{\mbox{${T}^{\prime}$}}
\newcommand{\Mprime}{\mbox{${M}^{\prime}$}}
\newcommand{\rprime}{\mbox{${r}^{\prime}$}}
\newcommand{\rfprime}{\mbox{${r}_f^{\prime}$}}
\newcommand{\vprime}{\mbox{${v}^{\prime}$}}
\newcommand{\tprime}{\mbox{${t}^{\prime}$}}
\newcommand{\mprime}{\mbox{${m}^{\prime}$}}
\newcommand{\Me}{\mbox{${M}_{e}$}}
\newcommand{\nh}{\mbox{${n}_{H}$}}
\newcommand{\rr}{\mbox{${R}_{2}$}}
\newcommand{\rf}{\mbox{${R}_{1}$}}
\newcommand{\vtwo}{\mbox{${V}_{2}$}}
\newcommand{\vout}{\mbox{${V}_{1}$}}
\newcommand{\dshell}{\mbox{${{\rho}_{sh}}$}}
\newcommand{\dwind}{\mbox{${{\rho}_{w}}$}}
\newcommand{\dslow}{\mbox{${{\rho}_{s}}$}}
\newcommand{\dfast}{\mbox{${{\rho}_{f}}$}}
\newcommand{\vfast}{\mbox{${v}_{f}$}}
\newcommand{\vslow}{\mbox{${v}_{s}$}}
\newcommand{\cc}{\mbox{${\rm cm}^{-3}$}}
\newcommand{\apj}{\mbox{ApJ}}
\newcommand{\apjl}{\mbox{ApJL}}
\newcommand{\apjs}{\mbox{ApJS}}
\newcommand{\aj}{\mbox{AJ}}
\newcommand{\araa}{\mbox{ARAA}}
\newcommand{\nat}{\mbox{Nature}}
\newcommand{\aap}{\mbox{AA}}
\newcommand{\gca}{\mbox{GeCoA}}
\newcommand{\pasp}{\mbox{PASP}}
\newcommand{\mnras}{\mbox{MNRAS}}
\newcommand{\apss}{\mbox{ApSS}}